# Mini-LANNDD T40: A detector to measure the neutrino-argon cross section and the $\nu_e$ contamination in the off-axis NuMI beam

David B. Cline[a], Youngho Seo[a], and Franco Sergiampietri[a,b]

[a]Department of Physics and Astronomy, University of California, Los Angeles, California 90095-1547, USA

[b]INFN-Sezione di Pisa, via Livornese 1291, San Piero a Grado (PI), Italy

**Abstract**

We describe a preliminary study of a 40-ton liquid argon TPC based on the ICARUS method to use in the NuMI near region in line with the LANNDD project [1,2]. This reduced-scale detector, called "Mini-LANNDD T40", is designed for R&D purposes and systematic measures on its response. Safety concerns are a key issue, which will be discussed as well as a preliminary design of the detector. Adapted as a near or vertex detector in a neutrino beam, the Mini-LANNDD T40 is capable of observing the $\nu_e$ flux in the off-axis beam, a key to use for measuring $sin^2 2\theta_{13}$ in the future, and measuring the low energy neutrino-argon cross-section, an important piece of information for future long baseline experiments.

## I. Introduction

The first test on the ICARUS T600 detector **[3]** has shown that the liquid argon TPC technique is able to provide event imaging of bubble-chamber quality, with a few MeV energy threshold and with the possibility to extend the active mass up to several tens of thousands of tons. Immersing the liquid argon TPC in a magnetic field opens the possibility of discriminating the charge sign of particles crossing the detector and measuring the momentum of muons that escape from the active volume. In view of large active mass detectors, as described in the LANNDD project, it is reasonable to start the experimentation with a "reduced-scale" prototype, which should be configured to allow a series of experimental uses:

- calibrate its energy response with electron and hadron beams,
- use as a near detector in a long-base line neutrino beam,
- use as a neutrino interaction vertex detector in front of higher mass detectors, and
- form a group of researchers and engineers with expertise in the noble liquid detector technique (purification, cryogenics, thermal insulation, low-noise electronics, DAQ, magnetic field, safety).

The Mini-LANNDD T40 detector project, described in section **II**, represents a possible detector design that could answer to the above-mentioned requirements. Section **III** is dedicated to the physics program that could be performed with such a detector.

## II. The Mini-LANNDD T40 detector

The Mini-LANNDD T40 detector (see **Figure 1**) is configured with a cylindrical liquid argon active volume, with horizontal axis. A wire chamber (Items 1 and 2 in **Figure 1**), aligned on a vertical plane along the cylinder axis, splits the active volume into two drift spaces, on the left and on the right. The wire chamber is made of two mirror sets of wire planes, providing the 2-coordinate readout, independently on each drift space. The drift is activated by a uniform electric field directed orthogonal to the wire planes. The field uniformity is ensured by a set of properly shaped and voltage-biased field shaping electrodes (Item 3). High voltage is fed to two cathode planes (Item 4) through a HV feedthrough (Item 6). Wire chamber signals, induced by drifting ionization electrons, are sent to the outer front-end electronics, via a set of six signal feedthroughs (Item 5).

The liquid argon is contained in a vacuum insulated and liquid nitrogen ($LN_2$) cooled cryostat. The inner vessel (Item 7) is surrounded by a $LN_2$ filled jacket (Item 8). The evacuated outer vessel (Item 9) provides the thermal insulation for the cold bodies. Three frames (Item 10) welded along the outer vessel, are used a) as feet, b) as stiffening rings, and c) as supports for the low heat conduction suspension belts (Item 11) for the inner vessel. Special flanges (Item 12), with low Z windows, are mounted on the inner and outer front shells for the energy calibration with an electron beam.

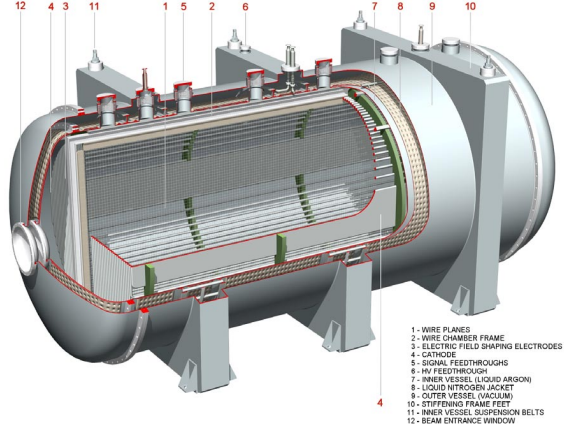

**Figure 1:** Cutaway view of the vacuum insulated, liquid $N_2$ cooled cryostat containing the liquid argon TPC.

The design parameters and the calculated performances of the detector are summarized in **Table 1**. Concerning the $LN_2$ consumption, we can note that vacuum insulation, eventually combined with super-insulation layers or with evacuated expanded perlite layers, can reduce the heat input $w_r$ through the walls at the level of fractions of *watt/$m^2$*. Purification of liquid argon, based on adiabatic closed-loop flow in liquid phase, does not require extra argon heating. Then the bulk of the heat input is due to the signal cables. If a deep economy on $LN_2$ is required, this can be further reduced by an optimized design of the heat exchanger.

The modest total heat input and $LN_2$ consumption reflect the stable and safe thermodynamic conditions in which the detector can be operated. Slow fluxes for nitrogen inside the heat exchanger and of the argon inside the purifier reduce the microphonic noise that can affect the *S/N* ratio on the read-out wires to the minimum. The good thermal insulation also minimizes the electric power requirements to a negligible level.

Embedding the liquid argon TPC in a magnetic field opens all series of experimental perspectives such as the possibility of identifying the charge sign of the particle initiating an electromagnetic shower and the momentum estimate of muons escaping from the active detector volume.

For the optimum detection of tracks bent in a magnetic field, the maximum bending plane should contain the drift direction. In case of operation on a particle beam, the magnetic field should be orthogonal to the beam axis. The mutual orientation of drift velocities, beam, and the magnetic field are sketched in **Figure 2**. While for the muon momentum measurement (over one meter or longer) a magnetic field $B \sim 0.3\,\mathrm{T}$ is adequate, for electron charge identification (over the first 2-3 radiation lengths), $B \sim 1\,\mathrm{T}$ is required.

The Mini-LANNDD T40, outfit with a dipole magnet, is represented in **Figure 3**. The cryostat design of Mini-LANNDD T40 can be realized as in **Figure 4**.

**Table 1:** Mini-LANNDD T40 - Parameters

| | |
|---|---|
| Total liquid argon volume | 42.8 $m^3$ |
| Gas argon volume | 0.9 $m^3$ |
| Active liquid argon volume | 27.7 $m^3$ |
| Active liquid argon sizes | $W$=2.4 $m$ ×$H$=2.3 $m$ ×$L$=6.0 $m$ |
| Active liquid argon mass | 38.5 *Ton* |
| Number of drift regions | 2 |
| Drift lengths | 2×1.2 m |
| Maximum required high voltage | 60 kV |
| Number of cathode planes | 2 |
| Number of HV feedthroughs | 1 |
| Number of wire chambers | 1 |
| Number of readout wire planes | 4 (6) |
| Orientation of readout wires | 0º , 90º |
| Number of readout wires | 5,632 |
| Number of signal feedthrough chimneys | 6 |
| Number of analog-to-digital processing crate pairs | 10 |
| Heat Input    a) Radiation (with $w_r$ = 1 *watt/$m^2$*) | 95 W |
|                b) Conduction (cables & mech. supports) | 210 W |
|                Total | 305 W |
|                Equivalent liquid nitrogen consumption | 0.2 $m^3$/d |

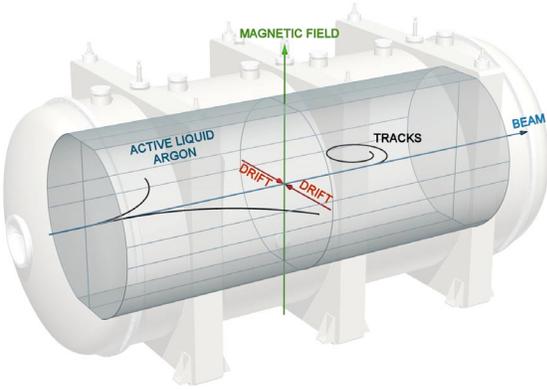

**Figure 2:** Mutual orientation of electric field (drift directions), magnetic field and beam axis for an optimum reconstruction of bent tracks.

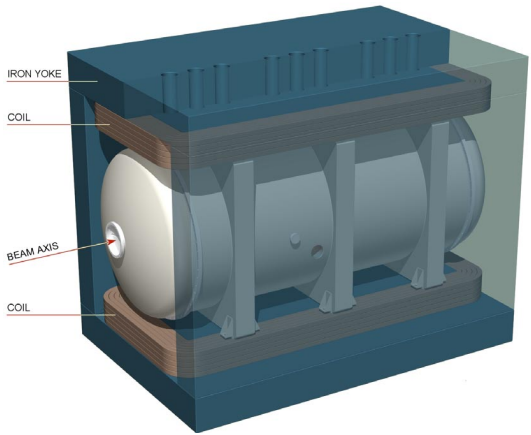

**Figure 3:** Representation of the Mini-LANNDD T40 detector inside a dipole bending magnet.

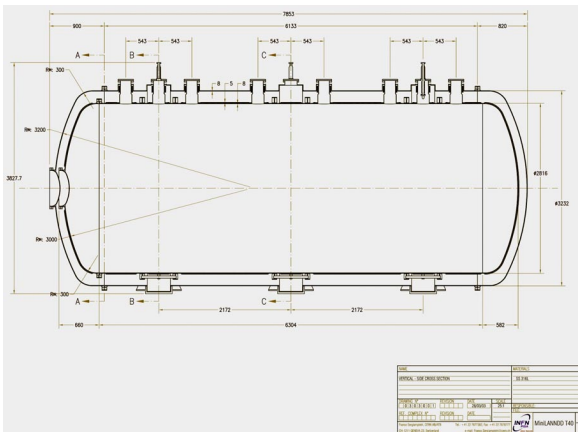

**Figure 4:** Details of Mini-LANNDD T40 cryostat.

### III. Physics

The physics potential of Mini-LANNDD T40 operating in the NuMI beam is (1) to measure the neutrino cross-sections on argon – this will be important for future long-baseline experiments such as ICARUS at the CNGS beam and a NuMI off-axis experiment, and (2) a possible measurement of the $\nu_e$ contamination in the NuMI beam or off-axis NuMI beam. **Figure 5** depicts the Mini-LANNDD T40 situated in the NuMi near hall in front of MINOS near detector.

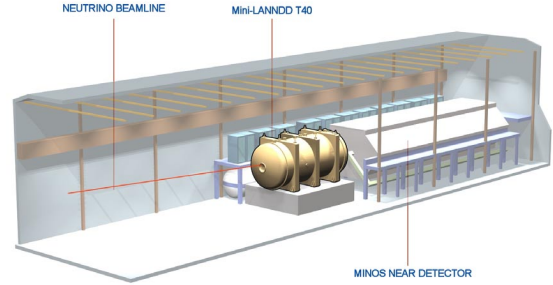

**Figure 5:** The Mini-LANNDD T40 detector situated inside the NuMI near hall (The drawing of the NuMI near hall is freely extracted from J. Morfin [6]).

### IV. Approximate detector cost estimate

We have made an estimate of the detector cost in the same manner as on estimates of the LANNDD and Mini-LANNDD (5-kT) cost [3]. **Table 2** gives this estimate. The use of Chicago Cyclotron superconducting magnet coil as discussed by K. McDonald [5] is not included for the estimate.

**Acknowledgements:** We wish to thank the ICARUS team for the development of this great detector. We also thank the Department of Energy for the support of the UCLA ICARUS team. Finally, we thank Kirk McDonald and Adam Para for discussions.

**Table 2:** Mini-LANNDD T40 – Rough cost estimate

| | | |
|---|---|---|
| Liquid argon | 60 $m^3$ | 30 k$ |
| Cryostat | Inner vessel<br>$LN_2$ jacket<br>Outer vessel<br>Beam entrance windows<br>Thermally decoupled chimneys for signals, HV, IN/OUT $Ar$, IN/OUT $LN_2$, inner vessel suspension | 500 k$ |
| Inner detector mechanics and wiring | Wire chamber (frame, wires, combs, and spacers)<br>Field shaping electrodes and cathodes | 70 k$ |
| Electronics (9,300 channels) and DAQ | Signal cables and feedthroughs<br>Analog and Digital Processing crates<br>Calibration pulser<br>Wire bias HV power supplies<br>Acquisition and event display computer | 350 k$ |
| Vacuum & cryogenic components | Vacuum pumps and gauges<br>$LN_2$ and $LAr$ storage dewars<br>$LAr$ purification system and purity monitor<br>Transfer lines, valves<br>Level and temperature monitors | 75 k$ |
| High voltage system | Power supply, feedthrough, monitor | 25 k$ |
| Other details & contingency | External trigger counters and electronics<br>UPS and $O_2$ monitors | 150 k$ |
| | Total | 1200 k$ |